\documentclass[10pt]{article}

\usepackage[margin=0.9in]{geometry}
\usepackage{amsmath}
\usepackage{amssymb}
\usepackage{graphicx}
\usepackage{booktabs}
\usepackage{array}
\usepackage{hyperref}
\usepackage{url}
\usepackage{setspace}
\usepackage{authblk}

\usepackage{tikz}
\usetikzlibrary{arrows.meta, shapes.geometric, decorations.pathreplacing, positioning, calc}

\definecolor{startcol}{RGB}{50, 50, 50}
\definecolor{decisioncol}{RGB}{230, 240, 255}
\definecolor{decisionborder}{RGB}{70, 100, 160}
\definecolor{endgreen}{RGB}{220, 245, 220}
\definecolor{endgreenborder}{RGB}{60, 140, 60}
\definecolor{endorange}{RGB}{255, 240, 220}
\definecolor{endorangeborder}{RGB}{200, 120, 40}
\definecolor{endblue}{RGB}{220, 235, 255}
\definecolor{endblueborder}{RGB}{50, 90, 170}
\definecolor{endgrey}{RGB}{240, 240, 240}
\definecolor{endgreyborder}{RGB}{130, 130, 130}
\definecolor{yesgreen}{RGB}{60, 140, 60}
\definecolor{nored}{RGB}{180, 40, 40}

\tikzset{
  terminal/.style={
    rectangle, rounded corners=6pt,
    draw=startcol, fill=gray!15,
    text width=4.2cm, align=center,
    font=\bfseries\small, inner sep=8pt
  },
  decision/.style={
    diamond, aspect=2.2,
    draw=decisionborder, fill=decisioncol,
    text width=3.8cm, align=center,
    font=\small, inner sep=3pt
  },
  outgreen/.style={
    rectangle, rounded corners=4pt,
    draw=endgreyborder, fill=endgrey,
    text width=4.6cm, align=center,
    font=\small, inner sep=7pt
  },
  outorange/.style={
    rectangle, rounded corners=4pt,
    draw=endgreyborder, fill=endgrey,
    text width=4.6cm, align=center,
    font=\small, inner sep=7pt
  },
  outblue/.style={
    rectangle, rounded corners=4pt,
    draw=endgreyborder, fill=endgrey,
    text width=4.6cm, align=center,
    font=\small, inner sep=7pt
  },
  outgrey/.style={
    rectangle, rounded corners=4pt,
    draw=endgreyborder, fill=endgrey,
    text width=4.6cm, align=center,
    font=\small, inner sep=7pt
  },
  arr/.style={-{Stealth[length=7pt]}, thick, draw=black},
  yeslabel/.style={font=\small\bfseries, text=yesgreen},
  nolabel/.style={font=\small\bfseries, text=nored},
}

\graphicspath{{Fig/}}

\hypersetup{
  colorlinks=true,
  linkcolor=blue,
  citecolor=blue,
  urlcolor=blue
}

\title{\textbf{Interpreting Net Survival: What We Estimate Versus What We Think We Estimate}}

\author[1]{Matthew J. Smith}

\affil[1]{Department of Medical Statistics, London School of Hygiene \& Tropical Medicine, London, United Kingdom, WC1E 7HT}

\date{}

\begin{document}

\maketitle

\begin{abstract}
Net survival is conventionally defined as ``survival if cancer were the only possible cause of death'', an estimand corresponding to cancer-specific mortality alone. The Pohar Perme estimator targets this by removing general population other-cause mortality from observed total mortality, but achieves it only when cancer patients experience the same other-cause mortality as the general population. However, cancer patients often experience elevated other-cause mortality due to baseline health differences and treatment-induced effects. Using recent theoretical work decomposing total mortality into four components (cancer deaths, baseline health differences, treatment-induced other-cause deaths, and general population other-cause mortality), we show that the Pohar Perme estimator delivers the sum of cancer deaths, baseline differences, and treatment-induced deaths, falling short of its intended estimand whenever either source of excess is present. From Botta \textit{et al}, we present empirical evidence showing relative risk of other-cause deaths ranging from 1.0 (colorectal cancer) to 4.0+ (head and neck cancers), and calculations demonstrating that net survival can substantially underestimate cancer-specific survival probability when relative risk exceeds 1.0. Critically, treatment-induced other-cause deaths represent irreducible causal pathways from cancer to death that cannot be eliminated through better stratification. We recommend interpreting net survival as ``survival where general population other-cause mortality is removed'' rather than as a causal counterfactual, and call for more precise language in cancer epidemiology.

\bigskip
\noindent\textbf{Keywords:} net survival, cancer-specific survival, relative survival, competing risks, causal inference, cancer epidemiology
\end{abstract}

\newpage


\section{Introduction}

Cancer survival statistics serve multiple purposes: monitoring population health, evaluating health systems, assessing treatment advances, and identifying disparities. To make meaningful comparisons across populations or over time, we must account for differences in background mortality unrelated to cancer. Net survival was developed to address this challenge by providing a ``pure'' measure of cancer survival, independent of other-cause mortality differences.

Net survival is conventionally interpreted as answering the question: ``What would survival be if cancer were the only possible cause of death?'' This interpretation appears in major international cancer survival studies, national cancer reports, and methodological tutorials.\cite{cronin2018} It is appealing because it suggests we have isolated the effect of cancer from all other competing causes of death.

However, this interpretation (while correctly stating the intended estimand) conflates what the measure aims to estimate with what it actually delivers. The Pohar Perme estimator attempts to recover this causal counterfactual by removing general population other-cause mortality rates from observed total mortality among cancer patients.\cite{perme2012} This operation achieves the intended estimand only under a specific condition: that cancer patients experience the same other-cause mortality as the general population. When this condition holds, removing general population mortality is sufficient to isolate cancer-specific mortality. When it does not hold, the estimator removes less than it needs to, and the result is not ``survival if cancer were the only cause of death'' but survival after removing only the general population component of other-cause mortality.

The problem arises when cancer patients experience elevated other-cause mortality relative to the general population. Modern cancer treatments can cause cardiovascular disease, infections, secondary malignancies, and other life-threatening complications, which are deaths causally attributable to having cancer, yet medically coded as other-cause deaths. Cancer patients may also differ from the general population in baseline health status, independently elevating other-cause mortality. The gap between what we think we are estimating and what we actually estimate has important implications for interpretation and use of net survival.

Recent work by Botta et al. provided a theoretical framework for understanding this issue by decomposing total mortality hazard into four components.\cite{botta2025} Building on their work, we clarify what the Pohar Perme estimator delivers, distinguish between addressable and irreducible sources of bias, present empirical evidence on the magnitude of the problem across cancer types, and provide recommendations for appropriate interpretation.

The remainder of this paper is structured as follows. In the Methods section, we present the hazard decomposition framework and introduce a directed acyclic graph that makes the causal structure of each component explicit, derive the role of relative risk as a diagnostic for interpretation failure, distinguish between the addressable and irreducible sources of excess other-cause mortality, and map all four relevant survival estimands to their appropriate research questions and audiences. In the Results section, we first review empirical evidence on variation in relative risk across cancer types, then use a historical prostate cancer trial to illustrate how the four components operate in practice, and finally present analytical calculations quantifying the divergence between net survival and the intended estimands across a range of relative risk values and follow-up times. We conclude with a discussion of implications for reporting practices and a call for more precise language in cancer epidemiology.


\section{Methods}

\subsection{Hazard Decomposition Framework}

The standard relative survival framework divides the total mortality hazard for cancer patients into two components: excess mortality (above what the general population experiences) and expected mortality (based on general population rates). Net survival is calculated by removing the expected component, leaving only excess mortality.

Pohar Perme et al. formalized this as:

\begin{equation}
h_{\text{total}}(t) = h_{\text{excess}}(t) + h_{\text{expected}}(t)
\end{equation}

where $h_{\text{expected}}(t)$ represents general population other-cause mortality and $h_{\text{excess}}(t)$ represents all mortality beyond this baseline. The Pohar Perme estimator targets survival in the hypothetical scenario where $h_{\text{expected}}(t)$ is removed.

Building on this framework, Botta et al. recently demonstrated that excess mortality itself comprises three distinct components, yielding a four-component decomposition \cite{botta2025}:

\begin{equation}
h_{\text{total}}(t) = h_A(t) + h_B(t) + h_C(t) + h_D(t)
\end{equation}

where:
\begin{itemize}
\item $h_A(t)$ is the cancer-specific mortality hazard (deaths from tumor progression, metastasis)
\item $h_B(t)$ represents baseline other-cause mortality differences between cancer patients and the general population, present at diagnosis and unrelated to cancer or its treatment
\item $h_C(t)$ represents treatment-induced other-cause mortality (deaths from cardiovascular disease, infections, or other conditions caused by cancer treatment)
\item $h_D(t)$ is the general population other-cause mortality hazard (equivalent to $h_{\text{expected}}$ above)
\end{itemize}

In this framework, the Pohar Perme estimator delivers the excess mortality hazard directly:
\begin{equation}
h_{\text{excess}}(t) = h_A(t) + h_B(t) + h_C(t)
\end{equation}

This decomposition clarifies that net survival equals cancer-specific survival ($h_A$ alone) only when both $h_B(t) = 0$ and $h_C(t) = 0$, that is, when cancer patients experience the same other-cause mortality as the general population.

This decomposition also clarifies the distinction between the net survival estimand, formally defined as ``survival if cancer were the only cause of death'' (Component A alone), and what the Pohar Perme estimator actually delivers. The estimator targets Component A by removing Component D, but as Equation 3 shows, this operation yields A+B+C whenever B or C are non-zero. The estimator performs its operation correctly; the problem is that the operation is insufficient to reach the intended estimand under realistic conditions.

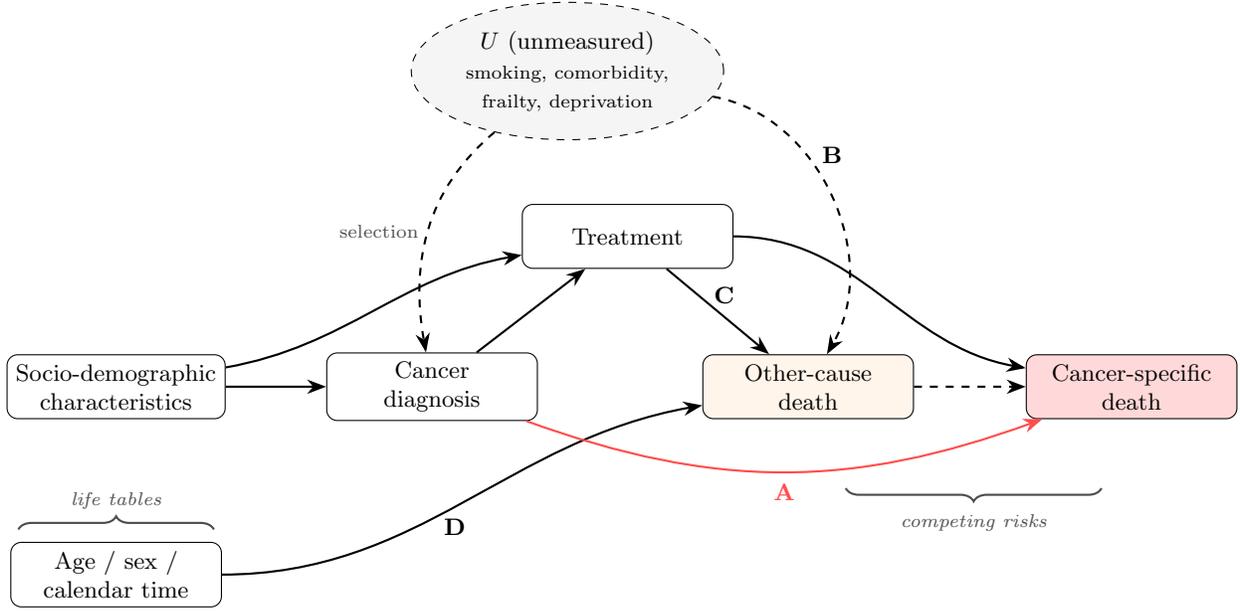
\begin{figure}[h]
\centering
\begin{tikzpicture}[
  arr/.style    = {-{Stealth[scale=1.1]}, thick},
  darr/.style   = {-{Stealth[scale=1.1]}, thick, dashed, black},
  box/.style    = {draw, rectangle, rounded corners=4pt,
                   minimum width=2.8cm, minimum height=0.85cm,
                   align=center, font=\small, fill=white},
  latent/.style = {draw, ellipse, dashed,
                   minimum width=3.6cm, minimum height=0.85cm,
                   align=center, font=\small, fill=gray!8},
  lbl/.style    = {font=\small\bfseries, inner sep=2pt}
]

\node[box]    (Age) at ( 0.0,  -1.0) {Age / sex /\\ calendar time};
\node[box]    (PatAge) at ( 0.0,  1.5) {Socio-demographic\\ characteristics};
\node[box]    (Cdx) at ( 4.2,  1.5) {Cancer\\ diagnosis};
\node[box]    (Tx)  at ( 6.8,  3.5) {Treatment};
\node[box, fill=orange!8] (Oc)  at ( 9.2,  1.5) {Other-cause\\ death};
\node[box, fill=red!15]   (Cs)  at (13.5,  1.5) {Cancer-specific\\ death};
\node[latent] (U)   at ( 6.0,  5.7)
  {$U$ (unmeasured)\\\scriptsize smoking, comorbidity,\\\scriptsize frailty, deprivation};

\draw[thick, black!70, decorate,
      decoration={brace, amplitude=5pt, mirror}]
  (9.7, 0.15) -- (13.1, 0.15)
  node[midway, below=6pt, font=\scriptsize\itshape, black!70]
  {competing risks};

\draw[thick, black!70, decorate,
      decoration={brace, amplitude=5pt, mirror}]
  (1.3, -0.4) -- (-1.3, -0.4)
  node[midway, below=-18pt, font=\scriptsize\itshape, black!70]
  {life tables};

\draw[arr] (Age) to[out=0, in=190]
  node[below, lbl, pos=0.48, yshift=-2pt] {\textbf{D}} (Oc);
\draw[arr] (PatAge) to[out=0, in=180] (Cdx);
\draw[arr] (PatAge) to[out=10, in=190] (Tx);
\draw[darr] (U) to[out=220, in=100]
  node[left, font=\scriptsize, text=black!70, pos=0.5] {selection} (Cdx);
\draw[darr] (U) to[out=350, in=60]
  node[right, lbl, pos=0.4, text=black, yshift=5pt] {\textbf{B}} (Oc);
\draw[arr] (Cdx) -- (Tx);
\draw[arr] (Tx) --
  node[right, lbl, pos=0.4, yshift=3pt] {\textbf{C}} (Oc);
\draw[arr] (Tx) to[out=0, in=170] (Cs);
\draw[darr, thick, black] (Oc) to[out=0, in=180] (Cs);
\draw[-{Stealth[scale=1.1]}, thick, red!70] (Cdx) to[out=-20, in=200]
  node[below, lbl, pos=0.5, yshift=-2pt, text=red!70] {\textbf{A}} (Cs);

\end{tikzpicture}
\caption{Directed acyclic graph of the four-component mortality hazard decomposition. Solid arrows represent direct causal effects; dashed arrows represent paths involving unmeasured confounding ($U$). The four main nodes lie on a common horizontal axis. \textbf{Component A} ($h_A$, red): cancer-specific deaths from tumour progression or metastasis (Cancer diagnosis $\rightarrow$ Cancer-specific death). \textbf{Component B} ($h_B$): excess other-cause mortality from unmeasured factors $U$ (smoking, comorbidity, frailty, deprivation) shared by cancer patients and the general population; a backdoor path in principle addressable by conditioning on proxies of $U$ through improved life table stratification. \textbf{Component C} ($h_C$): treatment-induced other-cause deaths along the causal chain Cancer diagnosis $\rightarrow$ Treatment $\rightarrow$ Other-cause death; irreducible because no stratification can remove a node on a causal path. Treatment may also influence cancer-specific death directly. \textbf{Component D} ($h_D$): background other-cause mortality determined by age, sex, and calendar time. The arrow from Other-cause death to Cancer-specific death reflects the competing risks structure: a patient who dies from another cause is no longer at risk of cancer-specific death, so other-cause death forecloses it. The lower brace indicates these two outcomes are competing events. The Pohar Perme estimator removes Component D via matched life tables but retains A, B, and C. Net survival therefore equals cancer-specific survival only when $h_B = h_C = 0$.}
\label{fig:dag}
\end{figure}

Figure~\ref{fig:dag} represents these four components as a directed acyclic graph, making explicit the causal pathways underlying each component and why Components B and C survive the net survival operation. The standard description of net survival (i.e., ``survival if cancer were the only cause of death'') implies switching off all pathways leading to other-cause death, leaving only Component A. In practice, the Pohar Perme estimator removes only Component D, closing just the path from age/sex/calendar time to other-cause death. Arrows B and C remain active, and the estimate of net survival therefore reflects A + B + C rather than A alone. The conventional interpretation holds only in the special case where B and C are both zero.

Crucially, we can observe empirically whether Components B and C are zero by examining other-cause mortality among cancer patients. If we isolate only the other-cause deaths (excluding Component A), the total other-cause mortality hazard among cancer patients is:

\begin{equation}
h_{\text{other-cause}}(t) = h_B(t) + h_C(t) + h_D(t)
\end{equation}

When $h_B(t) = 0$ and $h_C(t) = 0$, this equals the general population other-cause mortality $h_D(t)$. When Components B or C are present, other-cause mortality among cancer patients exceeds the general population rate. This excess provides the empirical signature that the standard interpretation of net survival has broken down.

\subsection{The Role of Relative Risk}

The relative risk (RR) of other-cause mortality quantifies the elevation in other-cause mortality among cancer patients relative to the general population. It emerges naturally from the hazard decomposition as the ratio of total observed other-cause mortality to expected general population rates:

\begin{equation}
\text{RR}(t) = \frac{h_B(t) + h_C(t) + h_D(t)}{h_D(t)}
\end{equation}

which simplifies to:

\begin{equation}
\text{RR}(t) = 1 + \frac{h_B(t) + h_C(t)}{h_D(t)}
\end{equation}

This reveals the direct connection between relative risk and the interpretation of net survival:

\begin{itemize}
\item \textbf{When RR $=$ 1.0:} The numerator equals the denominator, meaning $h_B(t) + h_C(t) = 0$. Components B and C are both zero. Cancer patients experience the same other-cause mortality as the general population. Net survival equals cancer-specific survival, and the standard interpretation is correct.

\item \textbf{When RR$>$1.0:} The numerator exceeds the denominator, meaning $h_B(t) + h_C(t) > 0$. Cancer patients experience elevated other-cause mortality beyond the general population. Net survival includes this excess (A + B + C) and does not equal cancer-specific survival (A alone). The standard interpretation breaks down.
\end{itemize}

Because the Pohar Perme estimator retains the additional hazard from Components B and C, the resulting net survival probability will be lower than disease-specific survival. The magnitude of RR directly indicates how much excess other-cause mortality is present, and therefore how far net survival departs from cancer-specific survival.

Botta et al. demonstrated that relative risk varies substantially across cancer types, ranging from approximately 1.0 for colorectal cancer to over 4.0 for head and neck cancers.\cite{botta2025} This variation indicates that the interpretation problem is not uniform but affects some cancer types far more than others. Moreover, because RR quantifies the combined magnitude of B + C, it provides a simple empirical indicator of when net survival departs from cancer-specific survival. However, RR alone cannot distinguish whether excess other-cause mortality arises from baseline health differences (Component B), treatment-induced effects (Component C), or both. As we discuss next, this distinction matters considerably: Component B is in principle addressable through better stratification, while Component C is not.

\subsection{The Distinction Between Components B and C}

While both Components B and C represent excess other-cause mortality among cancer patients, they differ fundamentally in their nature and implications:

\textbf{Component B (Baseline Differences):} These differences exist at diagnosis and reflect patient selection. Cancer patients may have different smoking rates, comorbidity burdens, socioeconomic status, or frailty than the general population. Importantly, Component B is \textit{addressable through methodological improvements}. More detailed stratification of life tables by these characteristics can reduce or eliminate B. Work is ongoing to develop stratified life tables accounting for deprivation, comorbidity, and other factors.\cite{rubio2021}

\textbf{Component C (Treatment-Induced Mortality):} These deaths result from a causal pathway: cancer diagnosis $\rightarrow$ treatment $\rightarrow$ other-cause death. A patient treated with anthracycline chemotherapy who later dies from heart failure has experienced a treatment-induced death. Critically, Component C is \textit{irreducible through methodological improvements}. No amount of life table stratification can remove C, because these deaths would not have occurred in the absence of cancer. They are causally attributable to cancer, even though medically coded otherwise.

This distinction matters for interpretation. When net survival includes substantial Component C, it is not merely a statistical artifact that better methods could address. Rather, it reflects a fundamental reality: cancer treatments save lives from cancer while sometimes causing deaths from other causes. These treatment-induced deaths are part of the burden of cancer.

Indeed, because Component C deaths would not have occurred but for the cancer diagnosis and its treatment, they are causally attributable to cancer and belong in any complete accounting of cancer's mortality burden. This observation motivates the distinction between disease-specific survival (Component A alone) and disease-attributable survival (Components A + C), which we develop in the next section.

\subsection{Mapping Estimands to Research Questions and Audiences}

Four estimands are relevant in cancer epidemiology, each designed to answer a distinct research question for a distinct audience. Understanding which estimand a given analysis produces, and whether it matches the intended research question, is essential for appropriate interpretation.

\textbf{Disease-specific survival (Component A only)} captures deaths from tumour growth, progression, and metastasis (i.e., deaths caused directly by the cancer). This is the estimand clinical trialists and oncologists primarily need when evaluating whether a treatment prevents cancer from progressing. A patient who dies from metastatic disease has experienced a disease-specific death; a patient who dies from treatment-induced heart failure has not, even though the cancer caused the treatment. In randomised controlled trials, disease-specific survival endpoints aim to estimate Component A, addressing the question: ``Does this treatment prevent death from cancer progression?'' This is the estimand of primary interest for regulatory approval.

\textbf{Disease-attributable survival (Components A + C)} captures all deaths causally attributable to having cancer, including both direct cancer deaths and treatment-induced other-cause deaths. A patient who dies from anthracycline-induced cardiomyopathy has experienced a disease-attributable death, even though the death certificate records a cardiovascular cause. These deaths would not have occurred but for the cancer diagnosis and its treatment, and they represent a real cost of cancer care. This estimand is most relevant to policy makers, health economists, and public health researchers asking: ``What is the complete mortality burden of cancer in our population?'' Cost-effectiveness analyses require disease-attributable survival because treatment-induced deaths must be counted against any survival benefit.

Table \ref{tab:scenarios} shows how net survival relates to these two clinical targets across different scenarios of component allocation.

\begin{table}[h]
\centering
\caption{What Net Survival Estimates Under Different Scenarios}
\small
\begin{tabular}{ccp{9.5cm}}
\hline
\textbf{Scenario} & \textbf{Net Survival Equals} & \textbf{Relation to Target Estimands} \\
\hline
None (B=0, C=0) & A & Equals disease-specific survival (ideal case) \\
\hline
Baseline only (B$\neq$0, C=0) & A + B & Underestimates disease-specific survival; gap equals Component B \\
\hline
Treatment only (B=0, C$\neq$0) & A + C & Underestimates disease-specific survival; equals disease-attributable \\
\hline
Both present (B$\neq$0, C$\neq$0) & A + B + C & Underestimates disease-specific survival (gap = B + C); underestimates disease-attributable survival (gap = B) \\
\hline
\end{tabular}
\label{tab:scenarios}
\end{table}

\textbf{Cause-specific survival (Components A + some C)} estimates cancer deaths as coded on death certificates, which typically includes Component A and some portion of Component C, depending on how physicians attribute treatment complications at the time of death. This estimand is most relevant to national cancer surveillance programs when cause-of-death data are reliable and consistently coded. It addresses the question: ``How many patients die with cancer listed as the underlying cause?'' Within a single country or region where coding practices are stable over time, cause-specific survival is often more interpretable than net survival because it does not require a population life table comparison.\cite{lambert2015}

\textbf{Net survival (Components A + B + C)} is the quantity delivered when the Pohar Perme estimator removes general population other-cause mortality from total observed mortality, retaining both baseline health differences and treatment-induced deaths. This estimand was developed specifically for international comparisons and settings where cause-of-death data are unreliable or coded inconsistently across countries. It addresses the question: ``How does cancer survival compare across populations with different background mortality rates?'' Major international studies such as CONCORD and EUROCARE appropriately use net survival for this purpose, as it enables valid comparisons even when cause-of-death coding varies between countries.\cite{allemani2018} However, it is precisely this retained excess (i.e., Components B and C) that prevents net survival from answering questions intended for disease-specific or disease-attributable survival. For this purpose (i.e., removing background mortality differences so that survival can be validly compared across populations with different life expectancies) A+B+C is the appropriate and intentional quantity. The Pohar Perme estimator correctly delivers what international comparisons require. The problem is not the measure but its label: describing net survival as `survival if cancer were the only cause of death' imports the interpretation of Component A alone, which A+B+C does not support when RR$>$1.

Table \ref{tab:estimands} summarises all four estimands with their component combinations, primary audiences, and appropriate use cases. The central interpretive problem with net survival is not the measure itself but its application outside its intended scope. Net survival has been routinely used to answer questions about disease-specific survival (Component A) or total cancer burden (Components A + C), when it actually estimates A + B + C. The discrepancy is largest when Component B is substantial, but even the irreducible Component C means that net survival cannot equal cancer-specific survival in settings where treatment causes other-cause deaths.

\begin{table}[t]
\centering
\caption{Mapping Survival Estimands to Research Questions and Audiences}
\footnotesize
\begin{tabular}{p{2.7cm}cp{6.5cm}p{4.5cm}}
\hline
\textbf{Estimand} & \textbf{Components} & \textbf{Interpretation} & \textbf{Primary Use Case} \\
\hline
Disease-specific survival & A & Survival if only the disease itself causes death (tumour progression, metastasis) & Clinical trials, treatment efficacy evaluation \\
\hline
Disease-attributable survival & A + C & Survival accounting for all deaths causally attributable to cancer (disease + treatment-induced) & Burden of disease studies, cost-effectiveness analyses \\
\hline
Cause-specific survival & A + (some C) & Survival based on cancer coded as underlying cause on death certificate & National surveillance when cause-of-death data reliable \\
\hline
Net survival & A + B + C & Survival with general population other-cause mortality removed & International comparisons, unreliable cause-of-death coding \\
\hline
\end{tabular}

\vspace{0.1cm}
\raggedright
\scriptsize
\textit{Note:} Components are defined as follows: A = cancer-specific mortality hazard (deaths from tumour progression, metastasis); B = baseline other-cause mortality differences between cancer patients and general population at diagnosis; C = treatment-induced other-cause mortality (deaths from cardiovascular disease, infections, or other conditions caused by cancer treatment); D = general population other-cause mortality hazard. Net survival is calculated by removing Component D from total observed mortality.
\label{tab:estimands}
\end{table}

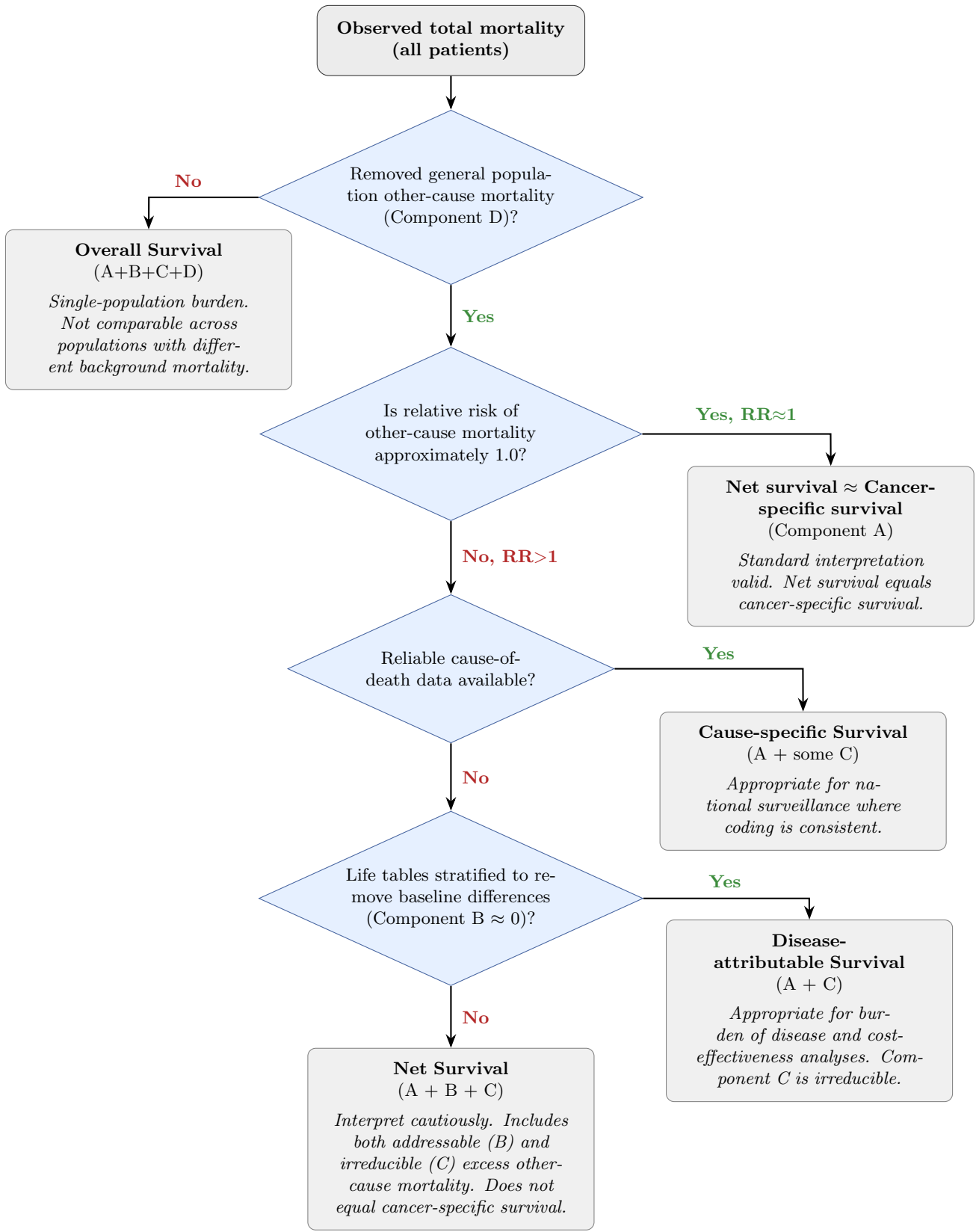
\begin{figure}[ht]
\centering
\begin{tikzpicture}[node distance=0.9cm and 2.6cm]

\node[terminal] (start) {Observed total mortality\\(all patients)};

\node[decision, below=0.6cm of start] (q1)
  {Removed general population other-cause mortality\\(Component D)?};

\node[outgrey, left=-0.6cm of q1, yshift=-2.0cm] (os)
  {\textbf{Overall Survival}\\(A+B+C+D)\\[4pt]
   \textit{Single-population burden. Not comparable across populations with different background mortality.}};

\node[decision, below=1.1cm of q1] (q2)
  {Is relative risk of other-cause mortality approximately 1.0?};

\node[outgreen, right=0.8cm of q2, yshift=-2.0cm] (cs)
  {\textbf{Net survival $\approx$ Cancer-specific survival}\\(Component A)\\[4pt]
   \textit{Standard interpretation valid. Net survival equals cancer-specific survival.}};

\node[decision, below=1.3cm of q2] (q3)
  {Reliable cause-of-death data available?};

\node[outblue, right=0.8cm of q3, yshift=-2.0cm] (css)
  {\textbf{Cause-specific Survival}\\(A + some C)\\[4pt]
   \textit{Appropriate for national surveillance where coding is consistent.}};

\node[decision, below=1.2cm of q3] (q4)
  {Life tables stratified to remove baseline differences\\(Component B $\approx$ 0)?};

\node[outgreen, right=0.4cm of q4, yshift=-2.0cm] (da)
  {\textbf{Disease-attributable Survival}\\(A + C)\\[4pt]
   \textit{Appropriate for burden of disease and cost-effectiveness analyses. Component C is irreducible.}};

\node[outorange, below=1.1cm of q4] (ns)
  {\textbf{Net Survival}\\(A + B + C)\\[4pt]
   \textit{Interpret cautiously. Includes both addressable (B) and irreducible (C) excess other-cause mortality. Does not equal cancer-specific survival.}};

\draw[arr] (start) -- (q1);
\draw[arr] (q1.west) -- +(-1.95,0) -- (os.north)
  node[nolabel, pos=0.09, above, yshift=0.1cm, xshift=0.7cm] {No};
\draw[arr] (q1.south) -- (q2.north)
  node[yeslabel, midway, right=2pt] {Yes};
\draw[arr] (q2.east) -- +(3.35,0) -- (cs.north)
  node[yeslabel, pos=0.08, above, yshift=0.1cm, xshift=-1.5cm] {Yes, RR$\approx$1};
\draw[arr] (q2.south) -- (q3.north)
  node[nolabel, midway, right=2pt] {No, RR$>$1};
\draw[arr] (q3.east) -- +(3.35,0) -- (css.north)
  node[yeslabel, pos=0.08, above, yshift=0.1cm, xshift=-1.5cm] {Yes};
\draw[arr] (q3.south) -- (q4.north)
  node[nolabel, midway, right=2pt] {No};
\draw[arr] (q4.east) -- +(2.95,0) -- (da.north)
  node[yeslabel, pos=0.08, above, yshift=0.1cm, xshift=-1.5cm] {Yes};
\draw[arr] (q4.south) -- (ns.north)
  node[nolabel, midway, right=2pt] {No};

\end{tikzpicture}
\caption{Decision tree for selecting the appropriate survival estimand.
Starting from observed total mortality, the researcher answers four sequential
questions to identify which estimand their analysis produces. Component letters correspond to the hazard decomposition in Equation~2.}
\label{fig:decisiontree}
\end{figure}

Figure \ref{fig:decisiontree} provides a decision tree that operationalises this estimand mapping for applied researchers. Starting from observed total mortality, the researcher answers four sequential questions: whether general population other-cause mortality has been removed; whether the relative risk of other-cause mortality is approximately 1.0; whether reliable cause-of-death data are available; and whether life tables have been sufficiently stratified to reduce Component B. Each branch leads to one of the four estimands described above, together with a brief statement of appropriate use. The tree makes explicit that net survival is the correct choice for international comparisons with unreliable cause-of-death coding, but that researchers should interpret it cautiously when RR substantially exceeds 1.0, and should consider disease-attributable survival when Component C is likely substantial.

\section{Empirical Evidence and Analytical Results}

\subsection{Variation in Relative Risk Across Cancer Types}

Empirical evidence from population-based cancer registry data demonstrates substantial variation in the relative risk of other-cause mortality across cancer types, depending on cancer site, age at diagnosis, sex, and time since diagnosis.\cite{botta2025} Using mixture cure models applied to EUROCARE-6 data, Botta et al. estimated RR across three cancer types (Table \ref{tab:rr_evidence}).

\begin{table}[h]
\centering
\caption{Estimated Relative Risk of Other-Cause Mortality by Cancer Type, Age Group, and Sex (Botta et al., 2025)}
\begin{tabular}{llccc}
\hline
\textbf{Cancer Type} & \textbf{Sex} & \multicolumn{3}{c}{\textbf{Age at Diagnosis}} \\
 & & \textbf{40--59} & \textbf{60--69} & \textbf{70--79} \\
\hline
Head and neck & Male   & 4.0 & 2.6 & 1.6 \\
              & Female & 4.5 & 2.9 & 1.8 \\
\hline
Breast        & Female & 1.3 & 1.4 & 1.2 \\
\hline
Colorectal    & Male   & \multicolumn{3}{c}{$\approx$1.0 across all age groups} \\
              & Female & \multicolumn{3}{c}{$\approx$1.0 across all age groups} \\
\hline
\end{tabular}
\label{tab:rr_evidence}
\end{table}

These differences have important implications. For colorectal cancer, RR is close to 1.0 across all ages and both sexes, meaning net survival closely approximates cancer-specific survival (i.e., Components B and C are minimal). For head and neck cancer, RR is substantially elevated, particularly in younger patients (RR up to 4.5 in females aged 40--59), reflecting the strong influence of shared risk factors such as smoking, alcohol, and HPV infection on other chronic diseases. Breast cancer represents an intermediate case, with modest RR elevation (1.2--1.4) that becomes clinically meaningful primarily in older patients and over longer follow-up.

Crucially, RR varies not only by cancer type but also by age and time since diagnosis. This means the interpretation problem is not static but worsens over time, particularly for cancer types and patient subgroups with elevated RR. Notably, for cancers such as head and neck, younger patients face a more severe interpretation problem than older patients because their RR is higher, even though their absolute background mortality ($h_D$) is lower.

Importantly, the source of RR elevation varies by cancer type. For head and neck cancer, the dominant contribution is likely Component B (shared risk factors such as smoking and alcohol elevate other-cause mortality at baseline, independent of treatment). For cancers treated with cardiotoxic regimens, Component C may contribute substantially. Because RR cannot distinguish between these two sources, and because they have fundamentally different implications (Component B being addressable through better stratification while Component C is not) understanding the composition of excess other-cause mortality matters as much as its magnitude. The following empirical example illustrates both components operating simultaneously.

\subsection{Empirical Decomposition: A Prostate Cancer Trial Example}

To illustrate how Components B and C operate in practice, we examined data from a randomized trial comparing high-dose estrogen therapy (5.0 mg diethylstilbestrol) to placebo in men with advanced prostate cancer, conducted between 1960 and 1975.\cite{vacurg1967} This trial provides an ideal opportunity to empirically decompose total other-cause mortality into its constituent components because: (1) treatment assignment was randomized, allowing clear attribution of treatment effects; (2) the era and patient characteristics allow matching to contemporaneous life tables; and (3) estrogen's cardiovascular toxicity provides a clear example of treatment-induced other-cause mortality.

The trial enrolled 125 patients assigned to high-dose estrogen and 127 assigned to placebo, with median follow-up of approximately 36 months. We calculated Component D (general population expected mortality) using age- and calendar year-matched life tables from the Human Mortality Database. Components B and C were estimated by comparing observed other-cause mortality in the placebo and estrogen arms to these expected rates.

Table \ref{tab:prostate_decomp} shows the decomposition. Component D was 13.5\%, representing expected other-cause mortality for men of this age over the trial follow-up period. The placebo arm experienced 45.7\% other-cause mortality, yielding Component B of 32.1 percentage points and a relative risk of 3.4 (45.7\%/13.5\%). This substantial baseline excess reflects the poor general health of men with advanced prostate cancer, that is, many had pre-existing cardiovascular disease, were heavy smokers, or had other comorbidities common in this population.

High-dose estrogen increased other-cause mortality to 52.8\%, adding Component C of 7.1 percentage points beyond the placebo arm and elevating the relative risk to 3.9 (52.8\%/13.5\%). This treatment-induced excess was primarily cardiovascular: 40.0\% of the estrogen arm died from cardiovascular causes compared to 28.3\% in placebo (difference: 11.7 percentage points). This cardiovascular toxicity of high-dose estrogen was well-documented historically and led to the abandonment of estrogen therapy for prostate cancer.

Component A (prostatic cancer deaths) comprised 29.1\% of the placebo arm and 21.6\% of the estrogen arm. However, this apparent 7.5 percentage point reduction in prostatic cancer mortality under estrogen does not represent a treatment benefit, as we discuss below.

\begin{table}[h]
\centering
\caption{Decomposition of Mortality at Median Follow-up: High-Dose Estrogen vs Placebo}
\small
\begin{tabular}{lccccc}
\hline
\textbf{Group} & \textbf{N} & \textbf{Prostatic Ca} & \textbf{Other-Cause} & \textbf{\% Other-} & \textbf{Relative} \\
 & & \textbf{Deaths (A)} & \textbf{Deaths} & \textbf{Cause} & \textbf{Risk} \\
\hline
General population (D) & --- & --- & --- & 13.5\% & 1.0 \\
Placebo (D + B) & 127 & 37 (29.1\%) & 58 & 45.7\% & 3.4 \\
High-dose estrogen (D + B + C) & 125 & 27 (21.6\%) & 66 & 52.8\% & 3.9 \\
\hline
\multicolumn{6}{l}{\small \textit{Component B (baseline differences): 32.1 percentage points}} \\
\multicolumn{6}{l}{\small \textit{Component C (treatment-induced): 7.1 percentage points}} \\
\hline
\end{tabular}
\label{tab:prostate_decomp}
\end{table}

This example demonstrates several key points. First, the relative risk substantially exceeds 1.0 in both arms (3.4 in placebo, 3.9 in estrogen), confirming that the standard interpretation of net survival as cancer-specific survival breaks down in this setting. Even in a randomized trial, Component B is substantial, representing 32.1 percentage points of excess other-cause mortality present at baseline. Second, Component C exists and is clinically meaningful: these treatment-induced deaths are causally attributable to cancer even when coded otherwise, and represent part of the mortality burden of cancer that better life table stratification cannot remove. Third, the Pohar Perme estimator applied to this trial would deliver Components A + B + C rather than its intended estimand (Component A alone). With B and C summing to 39.2 percentage points, the gap between what the estimator delivers and what it was designed to estimate is substantial and far exceeds the magnitude of any plausible treatment effect.

Examining net survival effects without this decomposition would be misleading. High-dose estrogen appeared to reduce prostatic cancer deaths (Component A) from 29.1\% to 21.6\%, a risk difference of approximately $-$7.5 percentage points suggesting benefit. However, this apparent protection occurs because estrogen kills patients through cardiovascular disease before they can die of prostate cancer, a classic example of Component C masking treatment harm through competing risks.\cite{young2020} The direct effect of estrogen on prostatic cancer progression (Component A alone) is likely null or even harmful, but this is obscured by the substantial mortality from Component C.\cite{young2020}

\subsubsection*{Interpreting the Trial Through Four Estimands}

The prostate cancer trial illustrates concretely how the choice of estimand changes the apparent conclusion about treatment benefit (Table \ref{tab:estimands}).

\textbf{Disease-specific survival (Component A only)} treats other-cause deaths as non-informative censoring. Under this estimand, estrogen appears beneficial: 21.6\% of the estrogen arm died of prostatic cancer versus 29.1\% in placebo, a difference of 7.5 percentage points suggesting meaningful treatment efficacy. A trialist reporting this estimand would conclude estrogen reduces cancer progression.

\textbf{Disease-attributable survival (Components A + C)} adds treatment-induced other-cause deaths back into the cancer burden. In the estrogen arm, Component C contributes 7.1 percentage points of treatment-induced cardiovascular mortality, yielding total disease-attributable mortality of approximately 28.7\% (21.6\% + 7.1\%). Compared to 29.1\% in the placebo arm, the apparent treatment benefit vanishes almost entirely. A health economist assessing the full burden of cancer treatment would correctly conclude estrogen offers no meaningful mortality advantage.

\textbf{Cause-specific survival (Components A + some C)} depends on how physicians coded the cardiovascular deaths attributable to estrogen therapy. If cardiovascular deaths were coded as unrelated to cancer (as was standard practice in this era) this estimand approximates Component A alone, again misleadingly suggesting treatment benefit. If some treatment-induced deaths were coded as cancer-related, the picture would improve, but inconsistently.

\textbf{Net survival (Components A + B + C)} removes only the general population hazard (Component D), retaining baseline health differences (Component B) in both arms equally. Since Component B is balanced by randomisation and is common to both arms, net survival in this within-trial comparison would show a similar pattern to disease-specific survival (again appearing to favour estrogen) while depressing the absolute survival values in both arms by the large Component B (32.1 percentage points). Net survival would therefore understate survival in both groups while still failing to reveal that the apparent treatment benefit is entirely attributable to Component C mortality displacing Component A mortality.

This illustrates a broader point: net survival's intended use case is international comparisons where cause-of-death data are unreliable. In a randomised trial with complete follow-up and known treatment assignment, there is no methodological justification for using net survival over disease-attributable survival and doing so obscures the very treatment harm that the trial was designed to detect.

This example illustrates why the choice of estimand is not merely a technical preference but determines the substantive conclusion: only disease-attributable survival correctly captures that estrogen therapy trades cancer deaths for cardiovascular deaths, providing no net survival benefit to patients.

\subsection{Analytical Results: Divergence of Net Survival from Intended Estimands}

To systematically quantify how Components B and C affect the divergence between net survival and its intended estimands, we derived survival probabilities analytically from the hazard decomposition. This approach yields exact results, since the divergence between net survival and cancer-specific survival is a mathematical consequence of the decomposition framework rather than a statistical phenomenon requiring simulation of individuals.

We used the following parameter specifications:
\begin{itemize}
\item Cancer-specific hazard: Weibull distribution with shape parameter 1.5 and scale parameter 5.3, yielding 40\% 5-year cancer-specific survival
\item General population other-cause hazard: constant rate $h_D = 0.025$ per year, approximating UK life table mortality for males aged 70 (i.e., representative of the age distribution in prostate, colorectal, and head and neck cancer populations)
\item Relative risk scenarios: 1.0, 1.5, 2.0, 3.0, 4.0
\item Component allocation: three scenarios: (1) all excess allocated to Component B (pure baseline differences); (2) all excess allocated to Component C (pure treatment effects); (3) equal split between B and C (mixed)
\end{itemize}

For each scenario and relative risk value, we derived three quantities. Disease-specific survival:
\begin{equation}
S_A(t) = \exp\bigl[-(t/5.3)^{1.5}\bigr]
\end{equation}
Disease-attributable survival:
\begin{equation}
S_{A+C}(t) = \exp\bigl[-(t/5.3)^{1.5} - h_C \cdot t\bigr]
\end{equation}
Net survival:
\begin{equation}
S_{\text{net}}(t) = \exp\bigl[-(t/5.3)^{1.5} - (\mathrm{RR}-1) \cdot h_D \cdot t\bigr]
\end{equation}

where $h_C = \text{frac}_C \times (\text{RR}-1) \times h_D$ and $\text{frac}_C$ is the proportion of excess hazard allocated to Component C. All calculations were implemented in R version 4.3.0.

\begin{figure}[h]
\centering
\includegraphics[width=\textwidth]{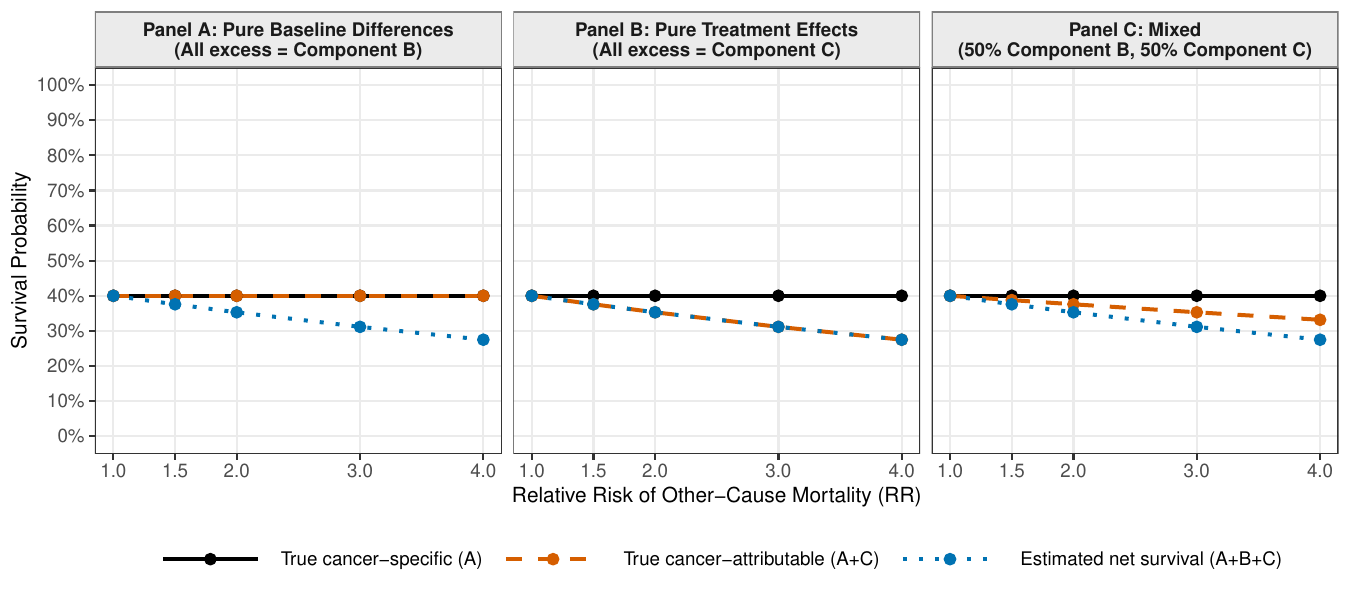}
\caption{Five-year survival probability by relative risk of other-cause mortality under three component allocation scenarios. Each panel shows disease-specific survival (Component A; black solid), disease-attributable survival (Components A + C; orange dashed), and estimated net survival (Components A + B + C; blue dotted) at five-year follow-up across relative risk values of 1.0, 1.5, 2.0, 3.0, and 4.0. Panel A: all excess other-cause mortality allocated to Component B (pure baseline differences). Panel B: all excess allocated to Component C (pure treatment effects). Panel C: equal split between B and C (mixed). Cancer-specific hazard: Weibull (shape 1.5, scale 5.3), yielding 40\% five-year cancer-specific survival. General population other-cause hazard: $h_D = 0.025$ per year (UK life tables, age 70). At RR = 1.0, all three estimands coincide.}
\label{fig:simulation_rr}
\end{figure}

Figure \ref{fig:simulation_rr} shows 5-year survival estimates across relative risk scenarios under the three component allocation schemes. At RR = 1.0, all three estimands coincide, confirming that net survival equals cancer-specific survival under the ideal conditions assumed by the standard interpretation. As RR increases, net survival diverges from both targets: at RR = 2.0, net survival understates cancer-specific survival by 4.7 percentage points at 5 years; at RR = 4.0, this gap reaches 12.5 percentage points.

\textbf{Scenario 1 (Pure Baseline Differences):} When all excess other-cause mortality represents baseline health differences (Component B), net survival underestimates both disease-specific survival and disease-attributable survival. At RR = 2.0, the gap between net survival and cancer-specific survival is 4.7 percentage points at 5 years; at RR = 4.0, this reaches 12.5 percentage points. This bias is addressable through better stratification. If we could perfectly match cancer patients to appropriate comparison groups in life tables, Component B would vanish and net survival would equal cancer-specific survival.

\textbf{Scenario 2 (Pure Treatment Effects):} When all excess represents treatment-induced mortality (Component C), net survival underestimates disease-specific survival but \textit{accurately estimates} disease-attributable survival. This is expected: Component C represents deaths causally attributable to cancer through treatment pathways, which disease-attributable survival correctly captures. The gap between net survival and cancer-specific survival is identical in magnitude to Scenario 1 for a given RR, since the total excess hazard is the same. Critically, this gap is \textit{irreducible}. No stratification can eliminate it because these deaths are caused by cancer treatment.

\textbf{Scenario 3 (Mixed):} In the realistic mixed scenario, net survival underestimates cancer-specific survival by the same amount as in Scenarios 1 and 2 for a given RR (since total excess hazard is unchanged), but now also underestimates disease-attributable survival by the Component B portion. Net survival therefore cleanly estimates neither target.

\begin{figure}[h]
\centering
\includegraphics[width=\textwidth]{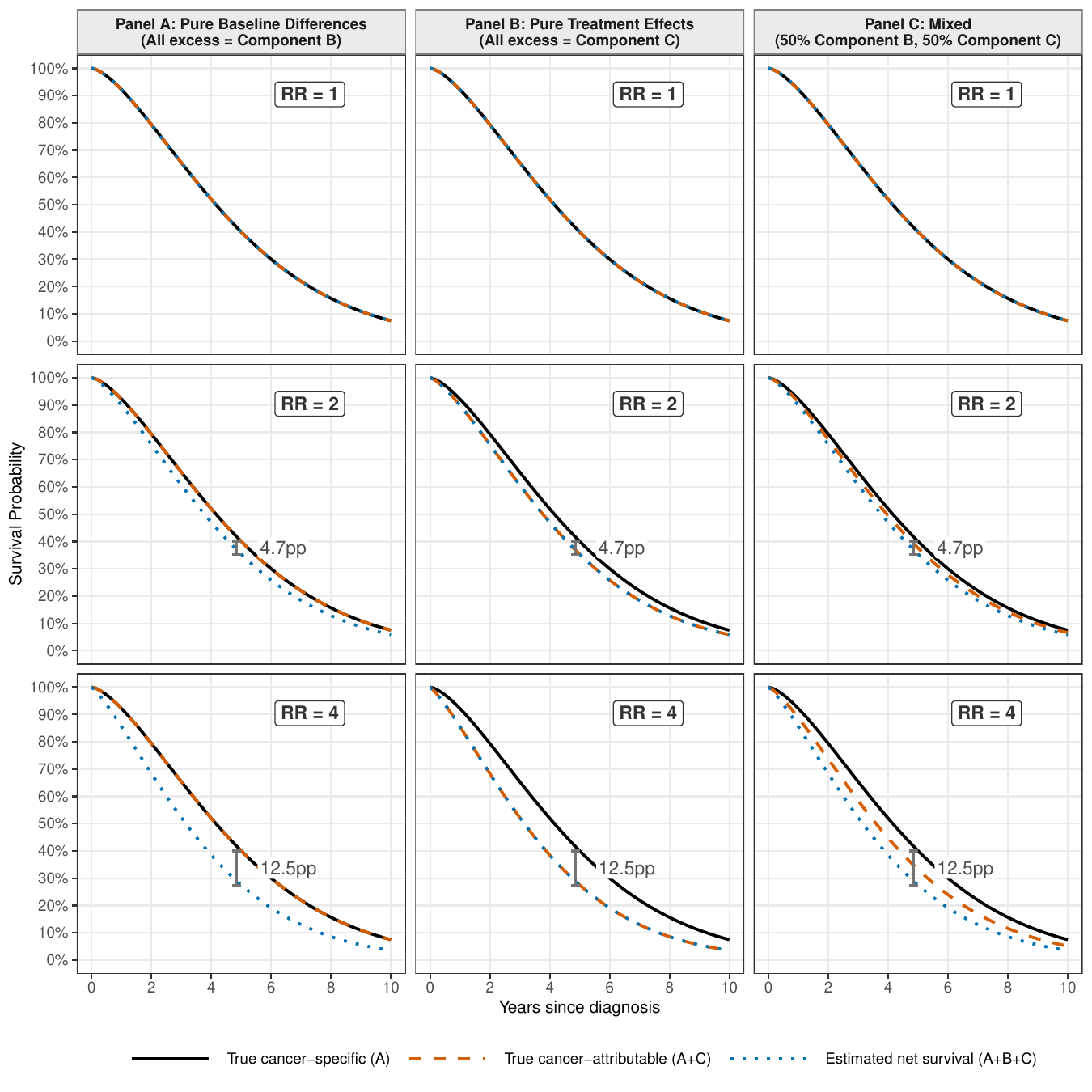}
\caption{Survival curves over 10 years by relative risk value and component allocation scenario. Panels are arranged in a $3 \times 3$ grid: rows correspond to relative risk values (RR = 1.0, 2.0, 4.0); columns correspond to component allocation scenarios (pure baseline differences, pure treatment effects, mixed). Within each panel, three lines show disease-specific survival (Component A; black solid), disease-attributable survival (Components A + C; orange dashed), and estimated net survival (Components A + B + C; blue dotted). Brackets at five years indicate the absolute gap (percentage points) between net survival and disease-specific survival. Cancer-specific hazard: Weibull (shape 1.5, scale 5.3). General population other-cause hazard: $h_D = 0.025$ per year. In RR = 1.0 panels, all three lines are coincident. The gap between net survival and cancer-specific survival widens from diagnosis and reaches a maximum around four to five years before narrowing as survival probabilities converge toward zero.}
\label{fig:simulation_time}
\end{figure}

Figure \ref{fig:simulation_time} shows survival curves over time at three fixed RR values (1.0, 2.0, 4.0) under each allocation scenario. This perspective reveals that the gap between net survival and cancer-specific survival is not constant over follow-up. It widens from diagnosis, reaches a maximum around 4--5 years (when 40\% of cancer-specific deaths have occurred and the cumulative impact of excess other-cause mortality is largest), and narrows toward 10 years as survival probabilities in all groups converge toward zero. At RR = 4.0, the maximum gap exceeds 12 percentage points, concentrated precisely in the medium-term window that is most clinically and policy-relevant. The RR = 1.0 panels confirm that all three estimands coincide when cancer patients experience no excess other-cause mortality.

\subsection{Implications of Analytical Results}

Three insights follow from these results. First, even modest RR elevation produces clinically meaningful divergence: at RR = 1.5, the gap is approximately 2.4 percentage points at 5 years, which is comparable to the survival improvements that cancer surveillance programs routinely report as progress. Second, RR alone cannot reveal whether excess other-cause mortality arises from Component B, Component C, or both, making it impossible to determine from net survival alone whether the gap is addressable or irreducible. Third, even ideally stratified life tables eliminating all Component B cannot recover disease-specific survival when Component C is present, because treatment-induced deaths are a genuine causal consequence of cancer, not a methodological artefact.

\section{Discussion}

\subsection{Principal Findings}

This analysis clarifies what net survival estimates and under what conditions it corresponds to cancer-specific survival. Using a theoretical framework that decomposes total mortality into four components, we distinguished two sources of excess other-cause mortality: baseline health differences (Component B) and treatment-induced deaths (Component C). Only the former is addressable through methodological improvements; the latter represents an irreducible causal pathway from cancer to death via treatment effects. Analytical results demonstrated that the magnitude of divergence can be substantial. With relative risk of 2.0 (i.e., a conservative estimate for many cancer types) net survival underestimates disease-specific survival by 4.7 percentage points at 5 years; at RR = 4.0, the gap reaches 12.5 percentage points. These are not minor technical discrepancies but have practical implications for interpreting survival trends, international comparisons, and treatment advances.

The prostate cancer trial illustrates this concretely: a 7.1 percentage point Component C was sufficient to completely mask the absence of any treatment benefit, and the correct conclusion (i.e., that estrogen trades cancer deaths for cardiovascular deaths) was only recoverable through disease-attributable survival. The current standard interpretation is therefore misleading in a specific and correctable sense. The phrase ``survival if cancer were the only possible cause of death'' correctly names the intended estimand (i.e., Component A alone) but the Pohar Perme estimator achieves this only under the restrictive condition that cancer patients experience the same other-cause mortality as the general population. The solution is not to abandon net survival but to replace a label that has systematically overpromised what the estimator delivers.

\subsection{Implications for Cancer Survival Research}

Our findings have several implications for how cancer survival statistics are reported and interpreted.

In terms of language and interpretation, we should stop describing net survival as ``survival if cancer were the only cause of death.'' This phrasing correctly names the intended estimand, but the Pohar Perme estimator does not deliver it when relative risk exceeds 1.0. Instead, we recommend: ``Net survival represents survival where general population other-cause mortality has been removed.'' This accurately describes the statistical operation without making unwarranted causal claims.

In terms of the choice of estimand, net survival remains valuable for certain purposes, particularly international comparisons where cause-of-death data may be unreliable or coded inconsistently. When cause-of-death data are reliable, cause-specific survival may be more interpretable because it acknowledges that treatment-induced other-cause deaths are causally attributable to cancer.

Regarding reporting practices, where feasible, researchers should report multiple measures: net survival (for continuity with existing literature and international comparisons), cause-specific survival (for interpretability), and standardized mortality ratios for other causes (to assess magnitude of excess). This combination would allow readers to understand both the burden of cancer deaths and the burden of treatment-induced other-cause deaths.

The magnitude of the interpretation problem varies by cancer type.\cite{botta2025} For colorectal cancer (RR $\approx$ 1.0), net survival and cancer-specific survival are approximately equivalent. For head and neck cancer (RR up to 4.0--4.5), they may differ substantially, such as by more than 12 percentage points at 5 years under our parameter assumptions. Researchers should consider the likely magnitude of Components B and C when interpreting net survival for specific cancer types or treatment eras.

Programs like CONCORD and EUROCARE, which report net survival across countries and time periods, should clarify interpretation in their reports.\cite{allemani2018} They might consider reporting standardised mortality ratios for other causes alongside net survival to help readers assess when relative risk deviates meaningfully from 1.0.

\subsection{Relation to Existing Literature}

Previous work has noted potential interpretation issues with net survival. Belot et al.'s tutorial mentions that the hazard framework underlying net survival makes assumptions about independence, but does not emphasise the implications of relative risk exceeding 1.0.\cite{belot2019} Dickman and colleagues have discussed the distinction between net survival and cause-specific survival but without the formal hazard decomposition framework.\cite{dickman2004}

Botta et al. provided crucial empirical evidence on excess other-cause mortality and introduced the four-component decomposition framework.\cite{botta2025} Our contribution builds on their work by explicitly connecting the decomposition to net survival interpretation, distinguishing addressable from irreducible sources of bias, and quantifying the magnitude of divergence analytically.

We emphasise that our critique is not of net survival as a measure per se, but of its interpretation. Net survival serves important purposes and will continue to be valuable. Our goal is to clarify what it estimates so it can be properly interpreted and used.

\subsection{Limitations}

Our analytical results are illustrative rather than exhaustive. We used a simplified Weibull distribution for cancer-specific hazards and a constant approximation for general population other-cause mortality, rather than replicating the full complexity of real cancer registry data. The scenarios we examined (pure B, pure C, mixed) are stylized; in reality, the allocation between Components B and C likely varies by cancer type, treatment era, and patient characteristics.

We relied on published empirical evidence from Botta et al. rather than conducting our own analysis of registry data.\cite{botta2025} The prostate cancer trial decomposition does constitute original empirical analysis, but it draws on a single historical trial with a specific patient population and treatment era, limiting its generalisability. While the registry evidence clearly demonstrates that relative risk varies across cancer types and often exceeds 1.0, more detailed empirical work would be valuable to understand the time-varying nature of Components B and C and their variation across patient subgroups.

Our framework assumes a competing risks structure, which may not fully capture semi-competing risks scenarios or time-varying relationships between cancer and other-cause mortality. We did not address related methodological issues such as period versus cohort analysis, age-period-cohort models, or temporal changes in cause-of-death coding. For clarity, our analytical results assumed constant other-cause hazard rates; the conceptual relationship between net survival and its component estimands holds generally under time-varying hazards, as it follows directly from the hazard decomposition framework.

\section{Conclusion}

Net survival is a valuable tool for cancer survival comparisons, but its standard interpretation requires correction. The conventional statement that net survival represents ``survival if cancer were the only possible cause of death'' correctly names the intended estimand, but the Pohar Perme estimator achieves this only when cancer patients experience the same other-cause mortality as the general population, which is a condition that empirical evidence shows is often violated. When it is violated, the estimator delivers A + B + C rather than Component A alone, and the label misrepresents what has been estimated.

When cancer patients face elevated other-cause mortality due to baseline health differences or treatment-induced effects, net survival will underestimate cancer-specific survival probability, equivalently, overstating cancer mortality burden. The treatment-induced component represents an irreducible causal pathway that cannot be eliminated through better stratification, and these deaths are arguably cancer-attributable even when medically coded otherwise.

We call for more precise language in cancer epidemiology. Net survival should be described as ``survival where general population other-cause mortality has been removed,'' not as a causal counterfactual about eliminating all other causes of death except from the disease of interest. Researchers should consider the likely magnitude of excess other-cause mortality when interpreting net survival for specific cancer types and treatment eras, and should report complementary measures where feasible.

Proper interpretation of net survival will strengthen cancer survival research by aligning our language with what we actually estimate, acknowledging the real burden of treatment-induced mortality, and helping readers understand what survival statistics can and cannot tell us about the impact of cancer.

\section*{Conflict of Interest}
The author declares no conflicts of interest.

\end{document}